\documentclass[twocolumn,preprintnumbers,superscriptaddress,endnote,nofootinbib,aps,prd,floatfix]{revtex4}
\usepackage[utf8]{inputenc}
\usepackage{lipsum}
\usepackage{footmisc,enumerate}
\usepackage{multirow}
\usepackage{subfigure}
\usepackage{amsmath}
\usepackage{graphicx}
\usepackage{placeins}
\usepackage{xspace}
\usepackage{hyperref}
\usepackage{todonotes}
\usepackage{mathrsfs}
\usepackage{amssymb}
\usepackage{bbm}
\usepackage{mathtools}
\usepackage{dsfont}
\usepackage{siunitx}
\usepackage{physics}
\usepackage{enumitem}

\definecolor{darkspringgreen}{rgb}{0.09, 0.65, 0.27}
\hypersetup{colorlinks=true, citecolor=blue, urlcolor=blue, linkcolor=blue}
\usepackage[normalem]{ulem}

\newcommand{\diag}{\mathop{\mathrm{diag}}}

\begin{document}

\newcommand{\og}{\ensuremath{\tilde{O}_g}\xspace}
\newcommand{\ot}{\ensuremath{\tilde{O}_t}\xspace}

\preprint{}

\title{Phenomenology of GUT-inspired gauge-Higgs unification}

\begin{abstract}
We perform a detailed investigation of a Grand Unified Theory (GUT)-inspired theory of gauge-Higgs unification. Scanning the model's parameter space with adapted numerical techniques, we contrast the scenario's low energy limit with existing SM and collider search constraints. We discuss potential modifications of di-Higgs phenomenology at hadron colliders as sensitive probes of the gauge-like character of the Higgs self-interactions and find that for phenomenologically viable parameter choices modifications of the order of 20\% compared to the SM cross section can be expected. While these modifications are challenging to observe at the LHC, a future 100 TeV hadron collider might be able to constrain the scenario through more precise di-Higgs measurements. We point out alternative signatures that can be employed to constrain this model in the near future.
\end{abstract}

\author{Christoph Englert} \email{christoph.englert@glasgow.ac.uk}
\affiliation{SUPA, School of Physics \& Astronomy, University of Glasgow, Glasgow G12 8QQ, UK\\[0.1cm]}
\author{David J.~Miller} \email{david.j.miller@glasgow.ac.uk}
\affiliation{SUPA, School of Physics \& Astronomy, University of Glasgow, Glasgow G12 8QQ, UK\\[0.1cm]}
\author{Dumitru Dan Smaranda} \email{d.smaranda.1@research.gla.ac.uk}
\affiliation{SUPA, School of Physics \& Astronomy, University of Glasgow, Glasgow G12 8QQ, UK\\[0.1cm]}


\maketitle

\section{Introduction}
\label{sec:intro}
The search for new physics beyond the Standard Model (BSM) is one of the key challenges of the current particle physics programme. Searches for deviations from the SM at large energies, most prominently at the Large Hadron Collider (LHC), which could point us in the direction of a more fundamental theory of nature have not revealed any statistically significant non-SM effects so far. In turn, the agreement with the SM of a plethora of measurements carried out at the LHC has cemented the SM as a surprisingly accurate electroweak scale description of the theory that completes the SM in the UV.

A final state that is typically highlighted as particularly relevant for the nature of the electroweak scale is multi-Higgs production~\cite{Baur:2002rb,Baur:2003gpa,Dolan:2012rv}, which is effectively limited to the analysis of Higgs pair production at both the LHC and future hadron colliders~\cite{Plehn:2005nk,Papaefstathiou:2015paa}.
Generic effective field theory (EFT) deformations can impact the di-Higgs rate dramatically~\cite{Goertz:2014qta,Carvalho:2015ttv}. This raises the question of the expected size of multi-Higgs production in the light of Higgs potential and other constraints~(see e.g. \cite{DiLuzio:2017tfn}).
EFT by construction can only provide limited insight in this context, i.e. constraints are only relevant when they can be meaningfully matched to a more complete UV picture~\cite{Alonso:2013hga,Jenkins:2013wua,Jenkins:2013zja,Englert:2014cva,delAguila:2016zcb,deBlas:2017xtg}. Analyses of concrete two-Higgs doublet and (next-to-)minimal supersymmetric SM scenarios~\cite{Basler:2018dac,Huang:2019bcs,Babu:2018uik,Adhikary:2018ise,Baum:2019pqc} (see also \cite{Cao:2013si}) have shown that once the heavy mass scales are decoupled, the low energy effective theory quickly approaches the SM expectation in these theories. Similar conclusions can be drawn for non-doublet representations see e.g.~\cite{Chang:2017niy}, and singlet extensions, e.g.~\cite{Englert:2019eyl}.

In this work we take a different approach compared to traditional scalar Higgs sector extensions and consider theories with gauge-Higgs unification~\cite{Espinosa:1998re,Hall:2001zb,Burdman:2002se,Medina:2007hz,Hosotani:2004wv}. In such scenarios, the self-interactions of the Higgs boson are fundamentally gauge-like. As these scenarios are effective theories in their own right, we base our investigation of the low energy effective interactions on the well-motivated UV constraint of grand unification (for a recent review see \cite{Croon:2019kpe}). Concretely, we consider the $SO(11)$ hybrid Grand Unified Theory (GUT) model in $\text{D}=6$ of Ref.~\cite{Hosotani:2017edv}.\footnote{A $SU(6)$-based scenario was discussed in \cite{Lim:2007jv}, also demonstrating that proton decay can be avoided.} To scan the model's parameters quickly, reliably and efficiently, we employ differential evolution techniques~\cite{Storn1997,Ring1996OnTU} specifically tailored to finding phenomenologically viable parameter regions. While we apply our approach to the this concrete theory, our implementation can be straightforwardly extended to other BSM scenarios.\footnote{A different variant of evolutionary algorithms, namely genetic algorithms, have been employed in the exploration of viable string theory scenarios and the pMSSM in~\cite{Abel:2018ekz,Abel:2014xta}.}

This work is organised as follows. In Sec.~\ref{sec:model}, we give a brief overview of the scenario of Ref.~\cite{Hosotani:2017edv} to make this paper self-consistent. Here we also introduce the relevant UV parameters that determine the low-energy physics. In Sec.~\ref{sec:parameter}, we detail our scan methodology to connect the UV picture with concrete phenomenological implications at the TeV scale. Sec.~\ref{sec:dihiggs} details di-Higgs physics for viable parameter choices. On the basis of LHC (and FCC-hh) projections of di-Higgs measurements and our scan results, we identify exotic states that will allow us to directly constrain this scenario in the near future in Sec.~\ref{sec:exotics}. We offer conclusions in Sec.~\ref{sec:conc}.

\section{$SO(11)$ Gauge-Higgs unified GUT}
\label{sec:model}
\subsubsection*{Geometry}
The model of Refs.~\cite{Hosotani:2017edv,Hosotani:2017hmu} is formulated on a 6D space-time with hybrid compactification. Concretely, we are working with a generalised Randall-Sundrum metric~\cite{Randall:1999ee}
\begin{equation}
 ds^2 = e^{-2 \sigma (y)} (\eta_{\mu\nu} dx^\mu dx^\nu  + d w^2)  + dy^2,
\end{equation}
where $e^{-2 \sigma (y)}$ is the warp factor along the compact $y$ direction and $\eta_{\mu\nu} = \diag(-1 , +1, +1, +1)$ is the flat 4D Minkowski space-time metric. $w$ denotes the second compactified euclidean coordinate. The two compact directions are referred to as the electroweak (EW) coordinate \mbox{$y \in [0, L_5]$} and GUT coordinate $w \in [0, 2\pi R_6]$, respectively.
Identifying space-time points via a $\mathds{Z}_2$ transformation as $(x^\mu, y, w) \rightarrow (x^\mu, -y, -w)$, results in
an orbifold $\mathcal{M}_4 \times (T^2 / \mathds{Z}_2)$. This space-time supports 5D branes at the orbifold fixed points $y= 0, L_5 $  with an anti-de Sitter bulk characterised by a cosmological constant $\Lambda = -10 k^2$.

\begin{figure}[t!]
 \begin{center}
  \def\svgwidth{0.9\columnwidth}
%
\begingroup%
  \makeatletter%
  \providecommand\color[2][]{%
    \errmessage{(Inkscape) Color is used for the text in Inkscape, but the package 'color.sty' is not loaded}%
    \renewcommand\color[2][]{}%
  }%
  \providecommand\transparent[1]{%
    \errmessage{(Inkscape) Transparency is used (non-zero) for the text in Inkscape, but the package 'transparent.sty' is not loaded}%
    \renewcommand\transparent[1]{}%
  }%
  \providecommand\rotatebox[2]{#2}%
  \newcommand*\fsize{\dimexpr\f@size pt\relax}%
  \newcommand*\lineheight[1]{\fontsize{\fsize}{#1\fsize}\selectfont}%
  \ifx\svgwidth\undefined%
    \setlength{\unitlength}{48.19805416bp}%
    \ifx\svgscale\undefined%
      \relax%
    \else%
      \setlength{\unitlength}{\unitlength * \real{\svgscale}}%
    \fi%
  \else%
    \setlength{\unitlength}{\svgwidth}%
  \fi%
  \global\let\svgwidth\undefined%
  \global\let\svgscale\undefined%
  \makeatother%
  \begin{picture}(1,0.90852082)%
    \lineheight{1}%
    \setlength\tabcolsep{0pt}%
    \put(0,0){\includegraphics[width=\unitlength,page=1]{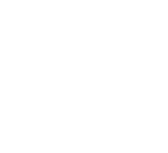}}%
    \put(0.09291316,0.7331412){\color[rgb]{0,0,0}\makebox(0,0)[lt]{\lineheight{0}\smash{\begin{tabular}[t]{l}\mbox{$\mathcal{M}_4\times S^1$}\end{tabular}}}}%
    \put(0,0){\includegraphics[width=\unitlength,page=2]{M4S1.pdf}}%
    \put(0.74589641,0.24375117){\color[rgb]{0,0,0}\makebox(0,0)[lt]{\lineheight{0}\smash{\begin{tabular}[t]{l}\mbox{$\mathcal{M}_4\times S^1$}\end{tabular}}}}%
    \put(0.18180201,0.5116609){\color[rgb]{1,0,0}\makebox(0,0)[lt]{\lineheight{0}\smash{\begin{tabular}[t]{l}\large\mbox{$\color{red} \langle\Phi\rangle$}\end{tabular}}}}%
    \put(0.00259922,0.23521795){\color[rgb]{0,0,0.50196078}\makebox(0,0)[lt]{\lineheight{0}\smash{\begin{tabular}[t]{l}\mbox{$SO(11)\color{red} \rightarrow SU(5)$}\end{tabular}}}}%
    \put(0.766261,0.49246437){\color[rgb]{0,0,0.50196078}\makebox(0,0)[lt]{\lineheight{0}\smash{\begin{tabular}[t]{l}\mbox{$G_\text{PS} \cap G_\text{SM}$}\end{tabular}}}}%
    \put(0.45110718,0.52604976){\color[rgb]{0,0,0.50196078}\makebox(0,0)[lt]{\lineheight{0}\smash{\begin{tabular}[t]{l}\mbox{$SO(11)$}\end{tabular}}}}%
    \put(0,0){\includegraphics[width=\unitlength,page=3]{M4S1.pdf}}%
    \put(0.14875143,0.0894403){\color[rgb]{0,0,0}\makebox(0,0)[lt]{\lineheight{0}\smash{\begin{tabular}[t]{l}\mbox{$0$}\end{tabular}}}}%
    \put(0.83491606,0.09235784){\color[rgb]{0,0,0}\makebox(0,0)[lt]{\lineheight{0}\smash{\begin{tabular}[t]{l}\mbox{$L_5$}\end{tabular}}}}%
    \put(0.4641158,0.18280452){\color[rgb]{0,0,0}\makebox(0,0)[lt]{\lineheight{0}\smash{\begin{tabular}[t]{l}\mbox{$y$}\end{tabular}}}}%
  \end{picture}%
\endgroup%

  \caption{$\mathcal{M}_4 \times (T^2 / \mathds{Z}_2)$ orbifold with 5D branes with a $\mathcal{M}_4 \times S^1$ topology at $y= 0, L_5$. $y$ corresponds to the warped coordinate, and the 5D branes have the extra dimensional flat coordinate corresponding to $w \in [0, R_6]$. The blue labels represent the $SO(11)$ symmetry in the 6D bulk, the same manifest UV brane symmetry, and with the effective Pati-Salam (PS) projection that results from the parity assignments intersection $G_\text{PS} \sim SO(6) \times SO(4) = SO(10) \cap SO(7) \times SO(4)$. The red labels represent the 5D $SO(11)$ spinor scalar field $\Phi_\mathbf{32}$ that breaks the UV brane symmetry down to $SU(5)$ via a Higgs mechanism, which in turn project the IR brane PS symmetry down to the SM.}
  \label{fig:pseudoRS}
 \end{center}
\end{figure}

Rewriting the metric in terms of the conformal coordinate $z  = e^{k y}$,
we have two associated mass scales
\begin{equation}
 m_{\text{KK}_5} = \frac{\pi k }{z_L - 1}   \qquad m_{\text{KK}_6} = \frac{1}{R_6},
\end{equation}
which are defined in terms of the first non-zero mass solution of the photon Kaluza-Klein (KK) tower with \mbox{$m_{\text{KK}_5} \sim \mathcal{O}(10)$ \SI{}{TeV}}, and the first non-zero mass mode along the GUT coordinate with $m_{\text{KK}_6} \sim \mathcal{O} ( M_\text{GUT})$.
The mass scales for the different fields are set by their parity assignments along either the EW or GUT dimension (Fig.~\ref{fig:pseudoRS}). Throughout this paper we will assume that there is a large
scale separation
$m_{\text{KK}_6} \gg m_{\text{KK}_5}$ (for a
qualitative approximation we set $M_\text{GUT} = 10^{16}$ \SI{}{GeV} in the following).

\subsubsection*{Matter content and interactions}
The matter content of the model consists of 6D and 5D fields.
The 6D matter fields are bulk fields and have a manifest $SO(11)$ gauge symmetry,
\begin{align*}
  \text{Gauge Bosons: }&\quad A_M(x, y, w) ,\\
  \text{Spinors: } &\quad \Psi_\mathbf{32}^{\alpha} (x, y, w) ,\\
  \text{Dirac Vectors: } &\quad \Psi_\mathbf{11}^\beta (x, y, w),~\Psi'^{\beta}_\mathbf{11}(x, y, w),
\end{align*}
where the $\mathbf{32}, \mathbf{11}$ subscripts represent the spinorial and vectorial representations of $SO(11)$, and $\alpha, \beta$ stand for generational indices, with $\alpha= 1, 2, 3, 4$, $\beta = 1, 2, 3$.

The 5D fields are confined to the UV brane at $y=0$, have a manifest $SO(11)$ gauge symmetry, and consist of:
\begin{align*}
    \text{Brane Spinor Scalar: }&\quad \Phi_\mathbf{32}(x, w) ,\\
    \text{Brane Symplectic Majorana  Spinor: }& \quad \chi_\mathbf{1}^\beta(x, w) ,
\end{align*}
where the $\mathbf{1}$ subscript stands for the singlet representation.

The matter fields come into effect via bulk and UV brane actions which have the general form
\begin{equation}
  S = \int d^{\,6}x \sqrt{-G} \left[\mathcal{L}_{\text{6D}} + \delta(y) \mathcal{L}_{\text{5D}} \right]\,,
\end{equation}
where $\sqrt{-G} = 1/(kz^6)$.
Starting of with the 6D Lagrangian, the gauge sector has the usual form for a Yang-Mills theory, accompanied by a gauge fixing term and ghost fields
\begin{equation}
 \mathcal{L}^\text{gauge}_\text{bulk} =  -\tr \left( \frac{1}{4} F^{MN} F_{MN} + \frac{1}{2\xi} (f_\text{gf})^2 + \mathcal{L}_\text{ghost}  \right).
\end{equation}
The bulk 6D action for the fermions is
\begin{multline}
 \mathcal{L}^\text{ferm}_\text{bulk} =  \sum_{\alpha=1}^{4} \overline{\Psi^\alpha_\mathbf{32}} \mathcal{D}(c_{\Psi^\alpha_\mathbf{32}})  \Psi^\alpha_\mathbf{32}  \\
 +  \sum_{\beta=1}^3 \overline{\Psi^{\beta}_\mathbf{11}} \mathcal{D}(c_{\Psi^{\beta}_\mathbf{11}})\Psi^{\beta}_\mathbf{11}
 +  \sum_{\beta=1}^3 \overline{\Psi'^{\beta}_\mathbf{11}} \mathcal{D}(c_{\Psi'^{\beta}_\mathbf{11}})\Psi'^{\beta}_\mathbf{11} \,,
\end{multline}
with bulk mass parameters $c_{\Psi^\alpha_\mathbf{32}}, c_{\Psi^\beta_\mathbf{11}}, c_{\Psi'^\beta_\mathbf{11}}$  for the fermions in their respective representation along with the generational index included in the covariant derivative definition (e.g.~\cite{Hosotani:2017edv}).

The brane-localised scalar in the spinorial representation $\Phi_\mathbf{32} (x, w)$ has a Higgs-like scalar potential
\begin{multline}
   \mathcal{L}_\text{brane}^\text{scalar} =
    -(D_\mu \Phi_\mathbf{32})^\dagger (D^\mu \Phi_\mathbf{32})
   {-(D_w \Phi_\mathbf{32})^\dagger (D_w \Phi_\mathbf{32}) }\nonumber\\
    -\lambda (\Phi_\mathbf{32}^\dagger \Phi_\mathbf{32} - |r|^2)^2\,.
\end{multline}
$\lambda, |r|$ determine the vacuum expectation value (VEV) that
$\Phi_\mathbf{32} (x, w)$ develops along the $SU(5)$ direction. This is then responsible for the breaking of the $SO(11)$ gauge symmetry on the UV brane.

On the same 5D brane, we have the brane symplectic Majorana fermions $\chi_\mathbf{1}^\beta(x, w)$, which facilitate the 6D seesaw mechanism~\cite{Hosotani:2017hmu} via
\begin{equation}
 \mathcal{L}_\text{brane}^\text{Maj.} = \frac{1}{2} \overline{\chi}_\mathbf{1}^\beta (\gamma^\mu \partial_\mu + \gamma^6\partial_w) \chi^\beta_\mathbf{1}
 - \frac{1}{2} M^{\beta\beta^{'}} \overline{\chi}_\mathbf{1}^\beta \chi_\mathbf{1}^{\beta^{'}} ,
\end{equation}
where $M^{\beta\beta^{'}}$ is a constant matrix.
Finally we have the Lagrangian terms that specify the coupling between the bulk 6D fermions and the 5D fields on the $SO(11)$ brane which induce effective Dirichlet  boundary conditions, and lift the mass degeneracy of the quark and lepton sector on the IR brane. The brane-localised action
contains eight allowed couplings between $\Phi_\mathbf{32}, \Psi_\mathbf{32} , \Psi_\mathbf{11}$ which are consistent with gauge symmetry, parity assignments  and keeping the action dimensionless.

\subsubsection*{Symmetry Breaking}
Symmetry breaking in this model consists of 3 stages which break $SO(11)$ down to $SU(3)_C \times U(1)_\text{EM}$ on the IR brane:
\begin{enumerate}[label={\arabic*)}]
 \item Symmetry breaking via orbifold parity assignments, which break $SO(11)$ to the Pati-Salam~\cite{Pati:1974yy} group $SO(11)\rightarrow SU(4)_C \times SU(2)_L \times SU(2)_R \equiv G_\text{PS}$ on the IR brane.
 \item Symmetry breaking via 5D brane interactions between the bulk gauge fields and $\langle \Phi_\mathbf{32} \rangle$, which break the $SO(11)$ symmetry down to $SU(5)$ on the UV brane.
 The zero mode spectrum on the IR brane has a SM symmetry content $ SU(5) \cap G_\text{PS} = G_\text{SM}$.
 \item Hosotani breaking~\cite{Hosotani:1983xw, Hosotani:1988bm, Hosotani:1983vn}, which acts as the electroweak symmetry breaking mechanism on the IR brane, breaking $G_\text{SM}$ to $ SU(3)_C \times U(1)_\text{EM}$ through a non-vanishing expectation value $\langle \theta_H \rangle$ of the associated Wilson loop. More specifically this happens through the $A_z$ component of the gauge field, which is a bi-doublet under the $SU(2)_L\times SU(2)_R$ and therefore plays the role of the usual SM Higgs boson~\cite{Hosotani:2009qf}.
\end{enumerate}

\subsubsection*{Effective Higgs Potential}

The equations of motion for the relevant towers, and how they relate to $SU(3)_C \times U(1)_\text{EM}$ via the twisted gauge imposed by the Hosotani mechanism, along with the computation of the effective potential is summarised in Appendix~\ref{appendix:EOMs}.

The free parameter set in charge of controlling the solution space consists of
\begin{equation}
\mathcal{P} = \left\{ k, z_L, c_0, c_1, c_2, c'_0, \mu_{1}, \tilde{\mu}_2, \mu_\textbf{11}, \mu'_\textbf{11}, M , m_B \right \}.
\end{equation}
$k$ is the $\text{AdS}_5$ curvature, $z_L$ the warp factor, $c_0, c_1, c_2, c'_0$ are the fermion bulk masses along the warped dimension $y$; $\mu_{1}, \tilde{\mu}_2, \mu_\textbf{11}, \mu'_\textbf{11}$ are couplings localised on the 5D UV brane (at $y=0$ in Fig.~\ref{fig:pseudoRS}) between the 5D scalar $\Phi_\mathbf{32}$ and the bulk fermion fields $\Psi^\alpha_\mathbf{32}, \Psi^\beta_\mathbf{11}, \Psi'^\beta_\mathbf{11}$, which have the effect of reducing the PS symmetry down to the SM on the IR brane.
Finally $M , m_B$ are 5D Majorana masses confined to the UV brane. All remaining parameters (see Sec.~\ref{sec:model}) are not relevant for the gauge boson and fermion equations of motion and, hence, do not impact our analysis.

The parameters determine the dynamical value of order parameter $\theta_H$ for electroweak symmetry breaking following the Hosotani mechanism.
The shape of the effective potential $V_\text{eff} (\theta_H)$ is sculpted by the bosonic and fermionic contributions. Following \cite{Hosotani:2017edv}, we focus on the 3rd generation, and identify $c_{\Psi^\alpha_\mathbf{32}} = c_0 , c_{\Psi^\beta_\mathbf{11}} = c_1, c_{\Psi'^\beta_\mathbf{11}} = c_2, c_{\Psi^4_\mathbf{32}} = c'_0$.
We have also set $m_B, M$ to the sample values stated by the authors in the original paper, $M = -10^7$ \SI{}{GeV}, $m_B = 1.145 \cdot 10^{12}$ \SI{}{GeV}, which is done to simplify the analysis and ensure the correct order of magnitude for neutrino masses (i.e. $<\SI{0.1}{eV}$).

The effective Higgs potential consists of the fermionic and bosonic contributions $V_\text{eff} (\theta_H) = V^\text{Bosons}_\text{eff} +  V^\text{Fermions}_\text{eff}$, arising from the relevant KK towers. For the explicit form of the contributions we refer the reader to the effective potential section in~\cite{Hosotani:2017edv}.
The mass of the Higgs boson is given by the second derivative of the effective potential
\begin{equation}
\label{eq:Hmass}
 m_H^2 = \frac{1}{f^2_H} \frac{d^2 V_\text{eff}(\theta_H) }{d \theta_H^2}  \eval_{\theta_H = \langle\theta_H\rangle},
\end{equation}
where
\begin{equation}
f_H =  \frac{\sqrt{6}}{(e / \sin \theta_W)} \frac{k}{ \sqrt{(1- z_L^{-1})(z_L^3 - 1)} }\,.
\end{equation}
Similarly, the trilinear coupling of the Higgs $\tau_H$, consists of the third derivative of the Higgs effective potential, which is then weighted by an appropriate power of $f_H$,
\begin{equation}
\label{eq:trilinear}
\tau_H = \frac{1}{6} \frac{1}{f_H^3} \frac{d^3 V_\text{eff} (\theta_H) }{d \theta_H^3} \eval_{\theta_H = \langle\theta_H\rangle}\,.
\end{equation}
Note that the Higgs potential is flat at tree level and is fully determined by the 1-loop radiative contributions.

\subsection{Consistent Parameter Regions}
\label{sec:parameter}
We now move on to the exploration of the model's low energy effective theory. This is done in a stochastic fashion, by randomly sampling the parameter space, finding the corresponding effective Higgs potential, and its minimum, which is then used to numerically solve the tower equations (appendix \ref{appendix:EOMs}).

In a first attempt to obtain phenomenologically viable parameter points, we uniformly random sample a parameter space point from our set of
input parameters, $\mathcal{P} = \{ p^i\}$, from within the corresponding bounds $\mathcal{P}_\text{bounds} = \{ (p^\text{min}_i, p^\text{max}_i) \}$ (i.e. $p^i \in [p^\text{min}_i, p^\text{max}_i]$). We then pass it through our coupling and mass spectrum computation to obtain the spectrum and relevant couplings and check its compatibility with SM constraints. Issues with uniform sampling along these lines arise when points require significant computation time only to find that they are in conflict with SM constraints and collider measurements. To reconcile this, at least in parts,
it turns out to be convenient to split the parameter set into two stages
\begin{gather*}
    \mathcal{P}_1 = \left\{ k, z_L \right\}\,, \quad \mathcal{P}_2  = \left\{c_0, c_1, c_2, c'_0, \mu_{1}, \tilde{\mu}_2, \mu_\textbf{11}, \mu'_\textbf{11} \right\}\,.
\end{gather*}
This choice enables us to pre-sample points, which directly reflect experimental constraints on the Kaluza Klein mass scale of \SI{4.1}{TeV}~\cite{Aaboud:2017yyg}
\begin{equation}
   m_{\text{KK}_5} = \frac{\pi k}{ z_L - 1} \geq \SI{4.1}{TeV}\,.
\end{equation}
The scan over the remaining parameters ${\cal{P}}_2$ is then performed within their respective boundaries.

In first instance, we define a set of general bounds
      \begin{gather*}
    \mathcal{P}_\text{bounds} =  \left \{  k \in [10^3 \SI{}{GeV}, 10^7  \SI{}{GeV}] , z_L \in [10, 2500], \right. \\
        \left. c_0 \in [0, 1], c'_0 \in [0, 1], c_1 \in [0, 2], c_2 \in [-3, 3]  \right . \\
    \left.  \mu_1 \in [0, 50] ,\tilde{\mu}_2 \in [0, 50] , \mu_\mathbf{11} \in [0, 50], \mu'_\mathbf{11} \in [0, 50]  \right\} .
   \end{gather*}
Similarly, we define the more restricted parameter range $\mathcal{P}^\text{extSol}_\text{bounds}$ which is obtained  by forming an appropriate extension from the sample solutions' parameters presented in Ref.~\cite{Hosotani:2017edv}
\begin{gather*}
   \mathcal{P}^\text{extSol}_\text{bounds} =  \left \{  k \in [ 10^5 \SI{}{GeV}, 5 \cdot 10^5  \SI{}{GeV}] , z_L \in [30, 60], \right. \\
   \left. c_0 \in [0, 0.8], c'_0 \in [0.1, 0.8], c_1 \in [0, 0.4], c_2 \in [-1.5, -0.2]  \right . \\
   \left.  \mu_1 \in [9, 15] ,\tilde{\mu}_2 \in [0, 3.5] , \mu_\mathbf{11} \in [0, 2.5], \mu'_\mathbf{11} \in [0, 2.5]  \right\}.
 \end{gather*}
In particular, we consider wide $z,k$ intervals. The latter criteria give rise to an adequate number of trial solutions. However, most of these are phenomenologically ruled out as they typically do not reproduce the SM mass
spectrum, predominantly due to $\langle \theta_H \rangle \simeq 0$ and periodic solutions. This behaviour is well-known from composite Higgs scenarios~\cite{Agashe:2004rs,Contino:2006qr,Contino:2003ve,Agashe:2006at,Ferretti:2016upr,DelDebbio:2017ini} (which are dual to the $\text{D}>4$ formulation in the sense of the AdS/CFT correspondence~\cite{Witten:1998qj,ArkaniHamed:2000ds,Rattazzi:2000hs}) where some fine-tuning is required to lift the Higgs mass and create a large mass gap between the electroweak scale and the UV composite scale. Yet, through the use of adapted techniques we can approach physically viable solutions for large ad-hoc parameter windows.

To identify the phenomenologically acceptable solutions we employ differential evolution~\cite{Storn1997,Ring1996OnTU} based on a global $\chi_G^2$ that parametrises the goodness of fit of the generated points given the experimental observations.
$\chi_G^2$ is defined as the unweighted sum of $\chi_i^2$ terms
   \begin{equation}
    \chi^2_G = \sum_{i\in \mathcal{C}} \chi^2_i \quad \text{with} \quad \chi^2_i \equiv \frac{(m_i - m_i^\text{Gen})^2}{( (\sigma_i^\text{Exp})^2 + (\sigma_i^\text{Th})^2)}
  \end{equation}
   where $\mathcal{C} = \{H, W^\pm, t, b, \tau \}$ for our purposes. $m_i$ is the central value of the masses being probed, $m_i^\text{Gen}$ is the generated mass given the parameter input, $\sigma_i^\text{Exp}$ are the experimental uncertainties. We also introduce a ``theoretical uncertainty'' $\sigma_i^\text{Th}$ of 1\%, see Tab.~\ref{tab:params}, to account for the RGE and threshold effects in the masses that we neglect. We also do not consider electroweak radiative corrections that affect input parameter relations. Both effects are usually small, see e.g. Refs.~\cite{Smaranda:2019opk, Athron_2017, Babu:2015bna, Hall:1993gn}.  We note that in the context of GUTs a special role is played by the Weinberg angle that we use as theoretical input to our scan (from which follows the $Z$ mass through SM relations).\footnote{We will explore the implications of the Weinberg angle and associated RGE effects in a forthcoming publication~\cite{toapp}.}

 \begin{table}[!t]
 \begin{tabular}{| c | c | c | c |}
 \hline
 state & mass $m$ [GeV] & $\sigma_H^\text{Exp}$ [GeV]  & $\sigma_H^{\text{Th}}$ [GeV]  \\
 \hline
 $H$ &  125.18 & 0.016 & 1.25 \\
 $W^\pm$ & 80.379 & 0.012 & 0.8037 \\
 $t$ &  172.44 &  0.9 & 1.724 \\
 $\tau$ & 1.776 & 0.00012 & 0.01776 \\
 $b$ & 4.18 & 0.04 & 0.0418  \\
 \hline
 \end{tabular}
 \caption{\label{tab:params} Parameter values for the definition of $\chi^2_G$. The experimental uncertainties are the most recent bounds~\cite{Tanabashi:2018oca} for the Higgs boson $H$~\cite{PhysRevD.98.030001}, the $W^\pm$ bosons, the top quark $t$, the bottom quark $b$, and the tau lepton $\tau$. We include a ``theoretical'' error to widen the parameter windows to discuss the phenomenological outcome in more detail below. The $Z$ mass is obtained through the Weinberg angle, which we use as an input. }
\end{table}

From the point of view of the infrared theory, in addition to the constrained SM masses, we need to reflect exclusion constraints from existing LHC searches that are relevant for the low energy spectrum of the model. As the most limiting searches, we include exotic quark searches~\cite{Aad:2015tba}, $Z'$ searches~\cite{Aaboud:2017sjh} as well as exotic charged lepton searches~\cite{ATLAS:2018ghc} to constrain the first non-SM KK states. By taking the aforementioned exclusion constraints at face value, if a parameter choice is conflict with any of these searches, we reject the point directly.

We deem a point as ``SM-like'' when its $\chi^2_G$ falls within the $95\%$ confidence limit bound for our degrees of freedom which selects a region
\begin{equation}
  \chi^2_G \leq 20.52 \label{eqn:LowChi2} \,.
\end{equation}

We can now consider $\chi^2_G$ as a cost function and look for points in the parameter space that minimise it. In addition, the $\chi^2_G$ evaluation can be time-consuming and can suffer from numerical singularities which makes the minimisation non-trivial. To more efficiently explore the parameter space, and find relevant solutions, we employ the  differential evolution algorithm
introduced by Storn and Price in~Ref.~\cite{Storn1997} (see also~\cite{Ring1996OnTU}). The algorithm uses the initial set of trial points described above to generate points that iteratively minimise the $\chi^2_G$ cost function.
The stochastic algorithm consists of four stages: initialisation,  mutation, recombination, and selection. It is designed as a parallelisable algorithm based on selection via a so-called ``greedy criterion''. Mutation, recombination and selection are then performed until we sufficiently minimise the cost function. Performing these routines is then referred to, as going through a generation, where we label the generation number with $G$.  We briefly outline the algorithm:
   \begin{itemize}
   \item  In the initialisation stage we randomly partition our initial population into subsets consisting of $N_P$ points. Each subpopulation is then treated separately, enabling (pseudo-)parallelisation of the algorithm. Each of the points has associated a $|\mathcal{P}| = 10$ dimensional parameter vector $\mathbf{p}^G$, formed by the corresponding point's parameter values.
   \item During the mutation stage we aim to generate a new parameter vector which will be used to generate a new point with smaller $\chi^2_G$ value. In this stage, we cycle through the points of the partition picking a random target point alongside three other distinct parameter points called ``donor points''. We label the target point parameter vector as $\mathbf{p}_t^{G}$, and the donor points parameter vectors as $\mathbf{p}_{d_1}, \mathbf{p}_{d_2}, \mathbf{p}_{d_3}$. From the 3 points we then form a ``mutation'' $\mathbf{p}_m^G$ by combining the parameter vectors,
      \begin{equation}
       \mathbf{p}_m^G = \mathbf{p}_{d_1}^G + F \cdot (\mathbf{p}_{d_2}^G - \mathbf{p}_{d_3}^G)
      \end{equation}
   where $F \in [0, 2]$ is a constant amplification factor to be set by the user.
   \item Recombination then aims to keep successfully minimised solutions of the current generation and improve on them by combining the target and the mutated points. The combination works as follows: To ensure that we have at least one component arising from the mutation vector we pick one parameter of the mutated vector $\mathbf{p}_m^G$ at random. The remaining parameters are adopted from the target vector $\mathbf{p}_t^{G}$, however we replace the $i$th component with the corresponding mutated entry with a uniform probability steered by a tunable decision factor $C_R$. This results in a combined parameter vector $\mathbf{p}_c^{G}$.
  \item In the last stage, selection, we compare the target $\mathbf{p}_t^G$ and the candidate point $\mathbf{p}_c^G$ by evaluating and comparing their respective cost function values. We admit to the new generation $G+1$ the point with the lowest cost function value. This is the admission via the ``greedy criterion''.
   \end{itemize}

   Mutation, recombination and selection are performed until we have treated all points within a generation as a target, which in turn determines the next generation. We keep iterating through generations until the cost function hits the threshold of a point being SM-like, specified by Eq.~\eqref{eqn:LowChi2}, or abort the process if no viable solution is obtained. This numerically minimises the cost function.

\begin{figure}
 \begin{center}
  \includegraphics[width = 1.0\columnwidth]{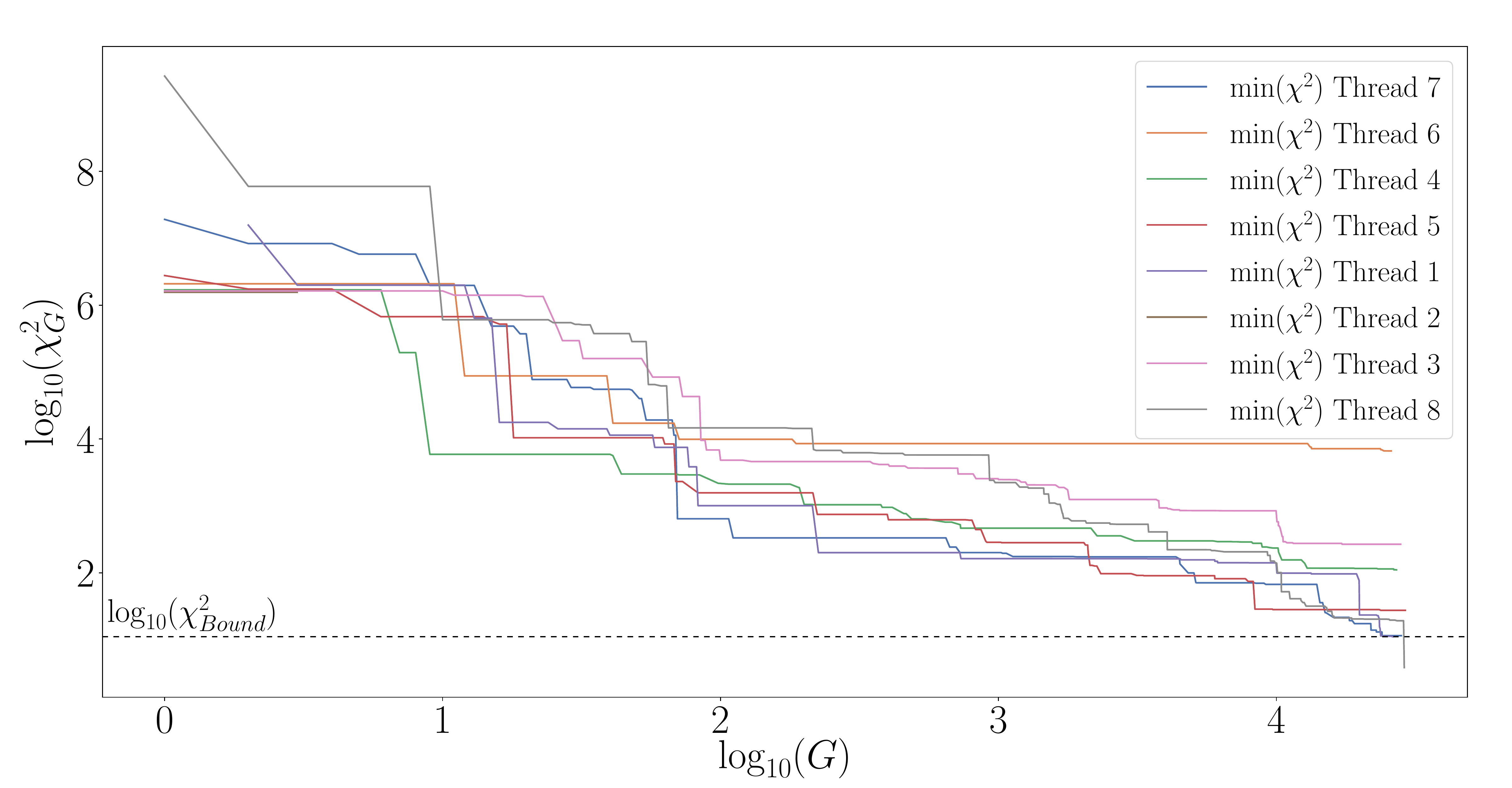}
  \caption{Log-log sample runs of the differential evolution outlined in the text, showing the $\chi_G^2$ value as a function of the generation number $G$. Note that each run (shown in different colours and denoted as ``thread'' specifies a parallel run) contains a population $N_P = 12$. The horizontal dotted line represents the $\log_{10}$ value of the SM like lower bound in Eq.~\eqref{eqn:LowChi2}, after which we terminate the thread. Note that this run was ended prematurely for Threads 5, 4, 3, 6, leading to non-SM solutions.}\label{fig:DiffEvolConvergence}
 \end{center}
\end{figure}

In obtaining results, the differential evolution parameters $N_P, F, C_R$ play important roles for convergence and its speed. Tuning $F, C_R$ to the problem at hand needs to be balanced against the population number $N_P$. By optimising these meta-parameters we can obtain adequate mutation and recombination rate which enables reliable convergence. For the extended parameter range $\mathcal{P}^\text{extSol}_\text{bounds}$, and method laid out in Sec.~\ref{sec:parameter}, the choices
\begin{equation}
  N_P = 12, \quad  C_R = 0.2368, \quad F = 0.6702,
\end{equation}
are appropriate (see also Ref.~\cite{pedersen2010good}).
We obtain  \mbox{$\chi_G^2\lesssim 20$} from an initial value of $\sim 10^{7}$ using on average $\sim 10^4$ generations (see Fig.~\ref{fig:DiffEvolConvergence}).

\subsection{Mass spectra}
\label{sec:spectra}

\begin{figure}[ht!]
\begin{center}
 \includegraphics[width = 1.0\columnwidth]{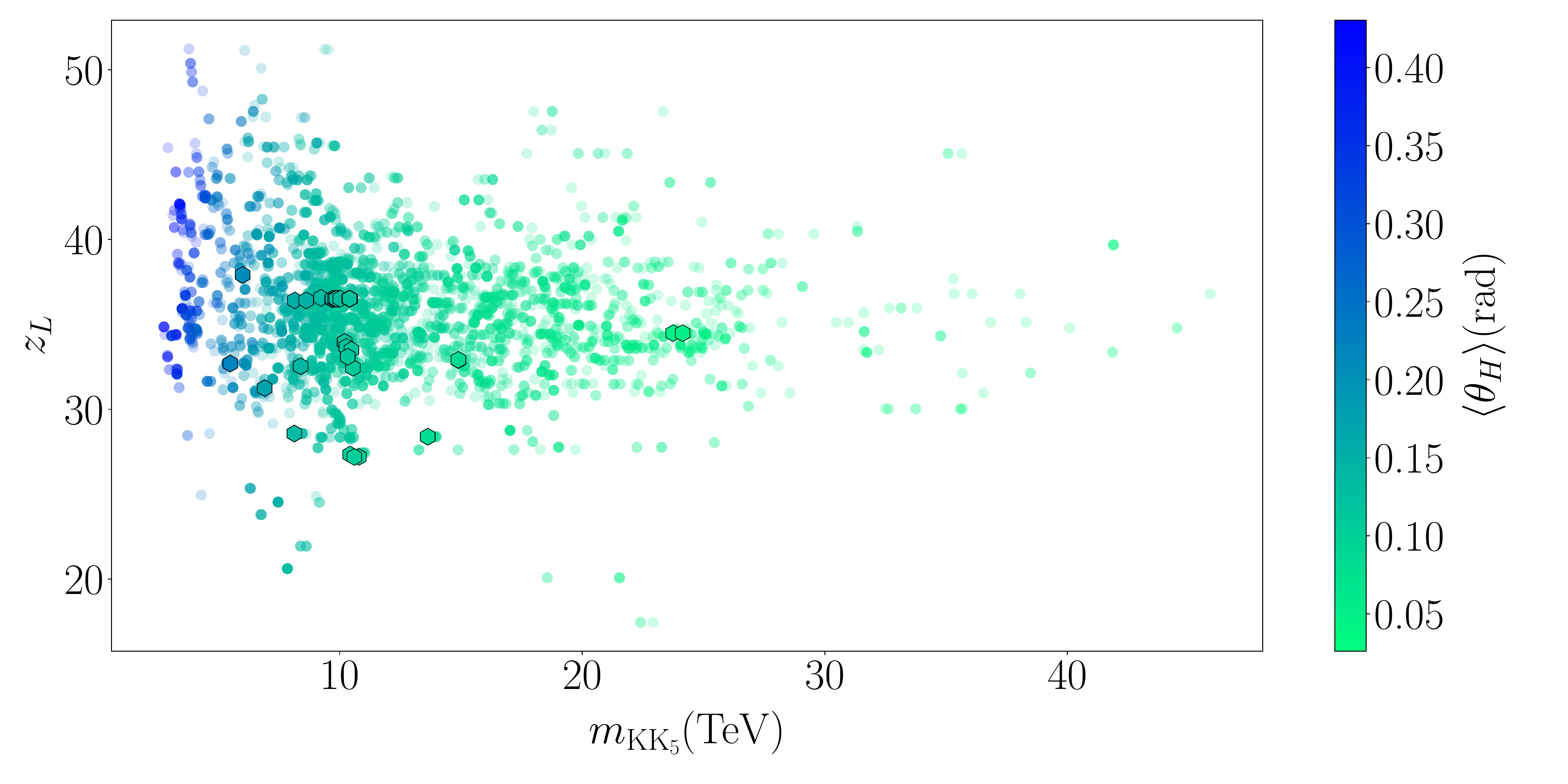}
 \caption{Scatter plot of representative parameter space points for the $SO(11)$ model before and after differential evolution as functions of the KK scale $m_{\text{KK}_5}$ and warp factor $z_L$. The color reflects the order parameter $\langle\theta_H\rangle$. Points highlighted as hexagons are the points that are SM-like (i.e. they obey the bound set in Eq.~\eqref{eqn:LowChi2}). Faded points are excluded on the basis of falling short of the $\chi^2_G$ measure bound.
 }\label{fig:mass}
\end{center}
\end{figure}

Employing the algorithm detailed in the previous section we can produce the consistent mass spectrum depicted show in Fig.~\ref{fig:mass}.
Direct LHC searches and our $\chi^2_G$ measure then reduce the viable solution space to the  points highlighted as hexagons in Fig.~\ref{fig:mass}, which serve as the basis of our discussion. From this we observe values of the order parameter $\langle\theta_H\rangle \lesssim  0.2$, which ensure a minimal deviations from the SM phenomenological values (see \cite{Funatsu:2017nfm}). Given that we require consistency with the observed Higgs mass, the theory cannot approach the decoupling limit. In other words,
the AdS/CFT dual of the symmetry-breaking Wilson loop becomes a Goldstone field if we send the UV cut-off to infinity. Therefore, a large mass gap between the KK scale and the Higgs mass is also not straightforward to achieve, which provides another motivation to implement the targeted numerical techniques detailed above. The differential evolution converges to solutions with a relatively low KK scale $M_{\text{KK}_5}$ for which the points are not yet excluded.

\section{Low energy phenomenology implications}
\label{sec:pheno}
\subsection*{Di-Higgs physics}
\label{sec:dihiggs}

We turn to the discussion of the low energy implications of the model that is now consistent with the SM mass spectrum. The implications for single Higgs physics (we denote the physical Higgs by $h$) have been discussed in Ref.~\cite{Funatsu:2013ni} (see also Ref.~\cite{Hosotani:2019cnn}), where it was shown that the model's single Higgs phenomenology is largely SM-like as a consequence of alternating contributions to the $H\to gg,\gamma\gamma$ decay (and production) loops. This is ultimately rooted in higher dimensional gauge invariance. Such a cancellation is broken in multi-Higgs final states and we therefore focus on this particular channel as a potentially sensitive probe of the model.

A recent projection by CMS~\cite{CMS:2018ccd} suggests that a sensitivity to $-0.18\leq \lambda^{95\%\,\text{CL}}_{\text{SM}}/\lambda_{\text{SM}}\leq 3.6 $ can be achieved, which corresponds to a gluon fusion cross section extraction of $0.85\leq\sigma(HH)/\sigma(HH)_{\text{SM}}\leq 2.39$ when assuming SM interactions. The inclusive SM di-Higgs cross section at the LHC is about $32$~fb~\cite{Dawson:1998py,Frederix:2014hta,deFlorian:2015moa,deFlorian:2016uhr,Borowka:2016ehy,Borowka:2016ypz,Heinrich:2017kxx,Grazzini:2018bsd,deFlorian:2018tah,Baglio:2018lrj}.
At a future FCC-hh, which is specifically motivated from a di-Higgs phenomenology perspective through the large inclusive cross section of 1.2~pb~\cite{Grazzini:2018bsd}, this could be improved to $\sigma(HH)/\sigma(HH)_{\text{SM}}\simeq [0.958,1.044]$, Ref.~\cite{Contino:2016spe} (see also \cite{Yao:2013ika,Barr:2014sga,He:2015spf,Papaefstathiou:2015iba,Adhikary:2017jtu,Banerjee:2018yxy,Banerjee:2019jys}).

\begin{figure}[!t]
 \begin{center}
  \includegraphics[width=1.0\columnwidth]{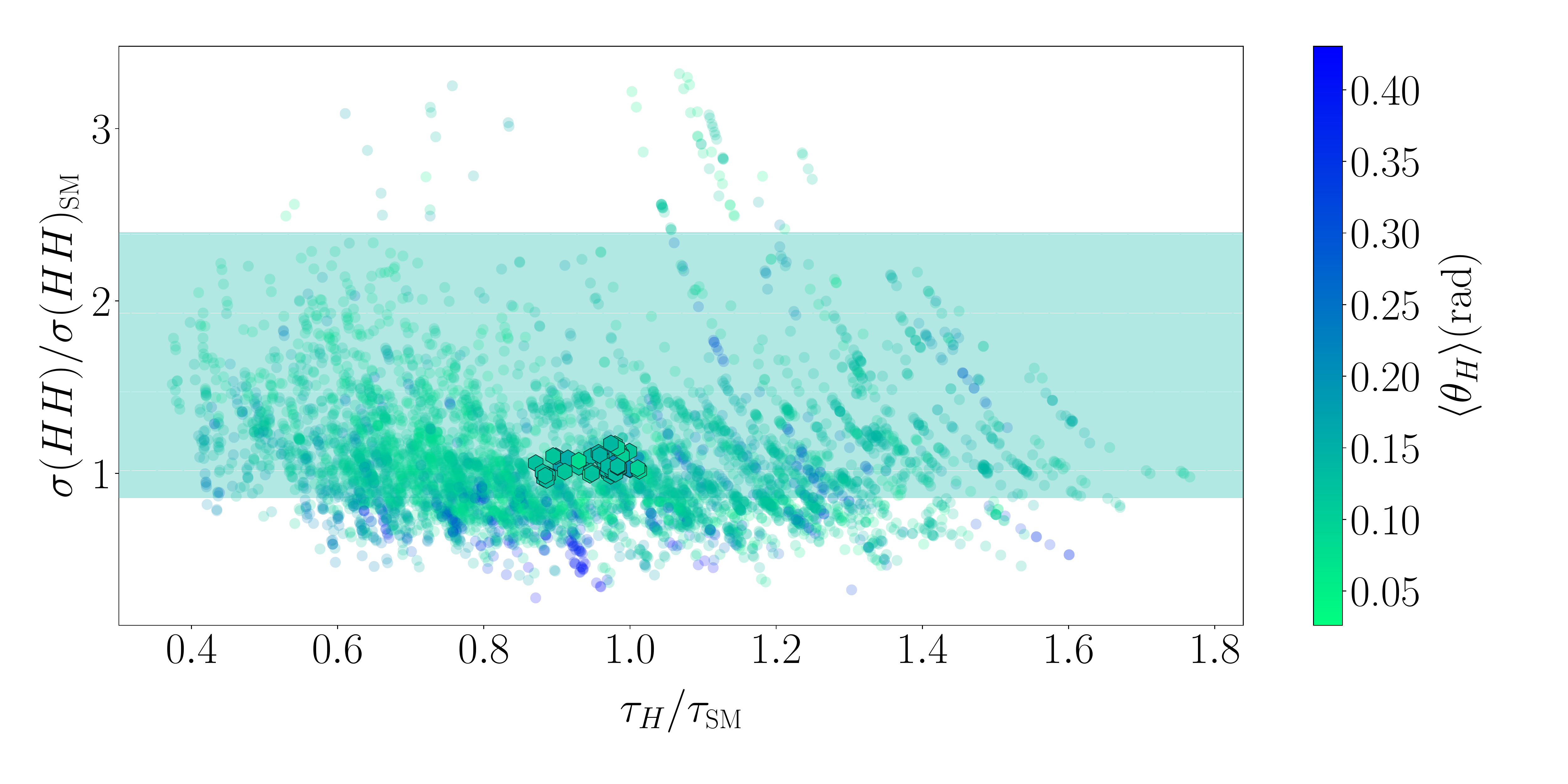}
  \caption{Scatter plot of representative parameter space points for the $SO(11)$ model before and after differential evolution as functions of the Higgs self-coupling relative to the SM $\tau_H/\tau_{\text{SM}}$ and di-Higgs cross section in relation to the SM. The color shading reflects the order parameter $\langle\theta_H\rangle$. Points highlighted as hexagons are the points that are SM-like (i.e. they obey the bound set in Eq.~\eqref{eqn:LowChi2}). The green band corresponds to the latest CMS di-Higgs measurement projection of Ref.~\cite{CMS:2018ccd}.}\label{fig:CMSsensitiv}
 \end{center}
\end{figure}
\begin{figure}[!b]
 \begin{center}
  \includegraphics[width=1.0\columnwidth]{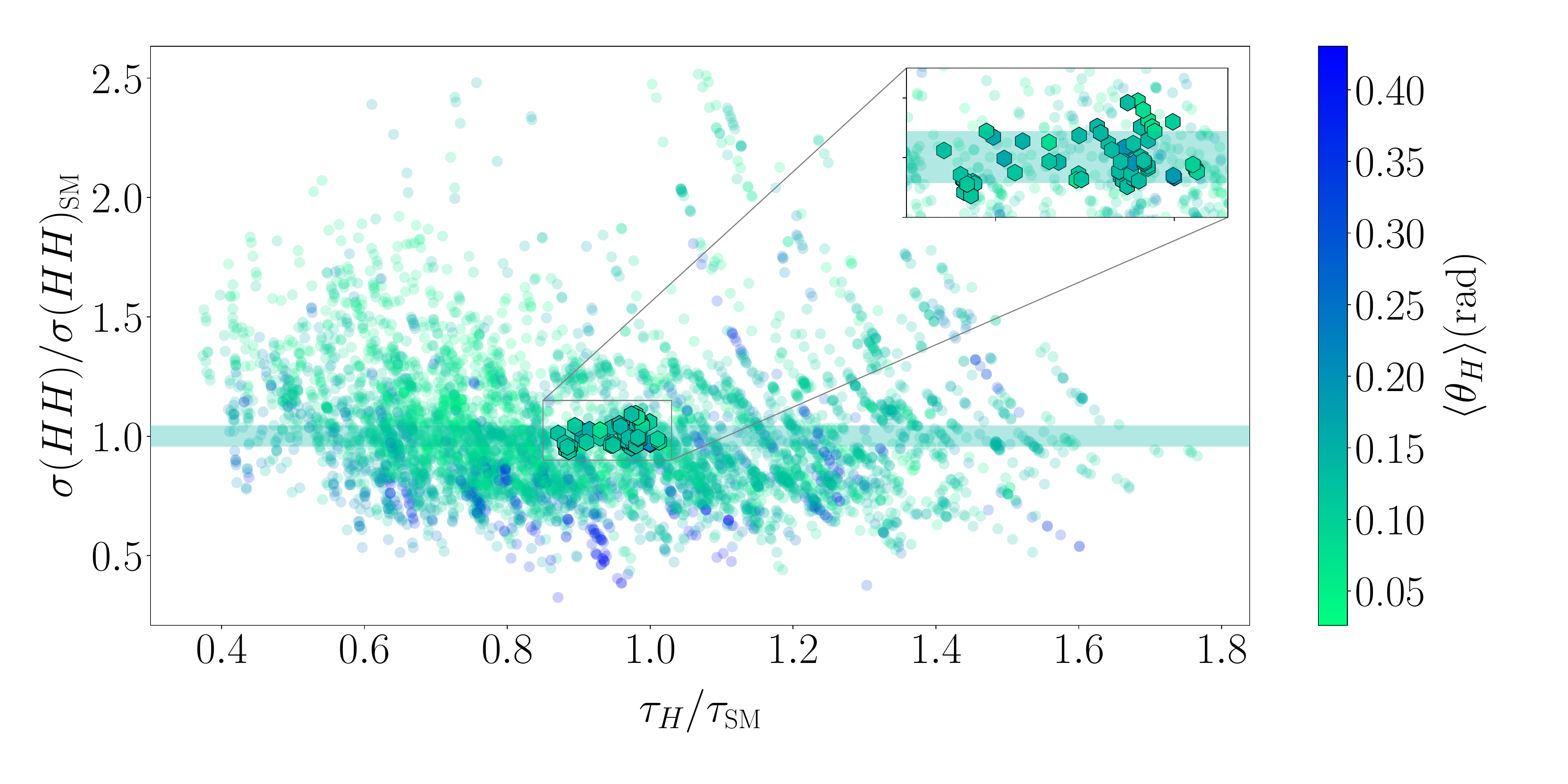}
  \caption{Scatter plot analogous to Fig.~\ref{fig:CMSsensitiv}, but for a 100 TeV FCC-hh. The green band now corresponds to the sensitivity region $0.958 \leq \sigma(HH)/\sigma(HH)_{\text{SM}}\leq 1.044$ which derives from a ${\cal{O}}(6\%)$ measurement of the Higgs self-coupling as detailed in \cite{Contino:2016spe}.}\label{fig:100TeVsensitiv}
 \end{center}
\end{figure}

Compared to the SM where the trilinear Higgs interaction is set by the Higgs vacuum expectation value and the Higgs mass, this correlation becomes modified in the present scenario. This extends to the top quark mass correlation with the vacuum expectation value, i.e. the top quark Yukawa coupling can be modified compared to the SM~\cite{Hosotani:2008by}. Both these effects are
relevant for di-Higgs production and we include them to a one-loop computation of $gg\to HH$ production~\cite{Glover:1987nx,Baur:2002rb,Dolan:2012rv}. We furthermore estimate the importance of the heavier states that arise in this scenario by means of the low energy effective theorem, but find that they do not significantly impact our result and their contribution is in the percent-range, below the expected theoretical uncertainty. In the following we will therefore focus on modifications of the cross section due to modifications away from SM parameters only.

\begin{figure}[!t]
 \begin{center}
  \includegraphics[width=1.0\columnwidth]{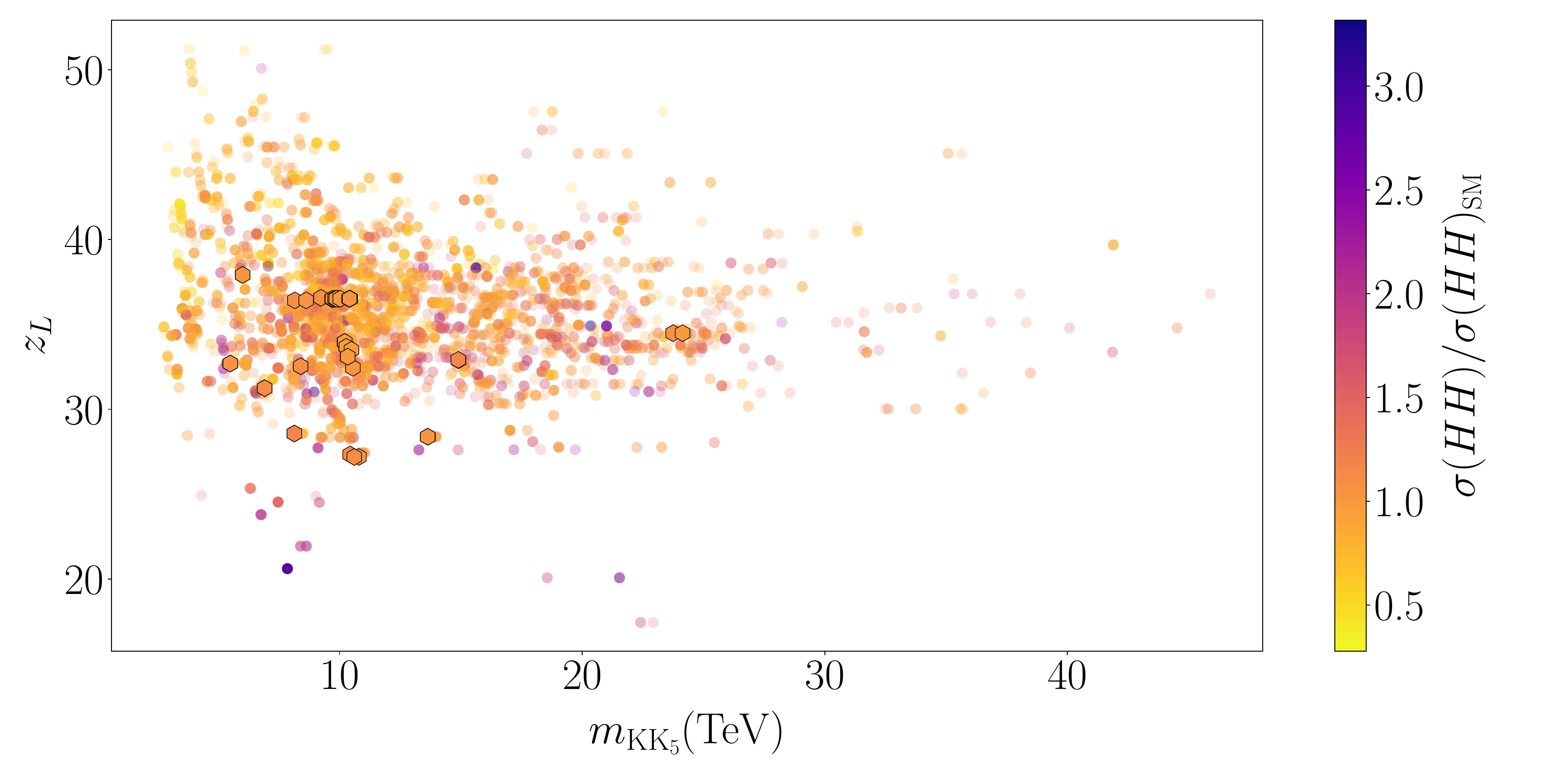}
  \caption{Scatter plot relating the KK scale with warp factor. The colour profile reflects the di-Higgs cross section modification away from the SM expectation. This is done to highlight the cross section ratio value for the points that obey the SM like bound set in Eq.~\eqref{eqn:LowChi2}\label{fig:xSectionRatio1000}. The highlighted hexagon points have $\sigma(HH) / \sigma(HH)_{\text{SM}}$ values within the interval $[0.961, 1.172]$, which is the envelope of cross section modifications that we observe. }
 \end{center}
\end{figure}

\begin{figure}[!b]
 \begin{center}
  \includegraphics[width=1.0\columnwidth]{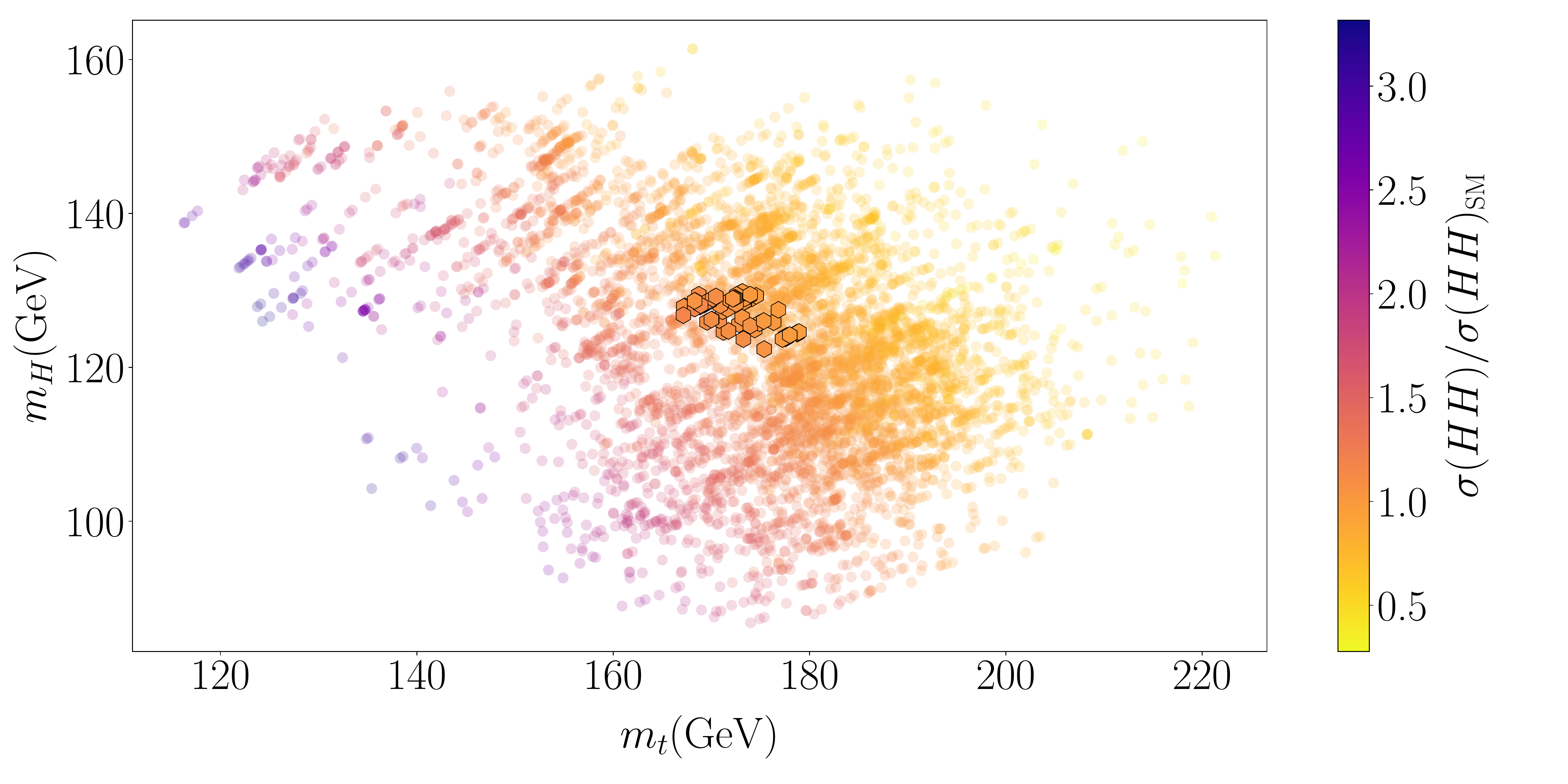}
  \caption{Scatter plot relating top quark and Higgs mass, with di-Higgs cross section modification at 13 TeV shown as colour shading, where the highlighted hexagon points have $\sigma(HH) / \sigma(HH)_{\text{SM}}$ values within the interval $[0.961, 1.172]$.}\label{fig:TopHiggsxSect}
 \end{center}
\end{figure}

The results are summarised in Figs.~\ref{fig:xSectionRatio1000} and \ref{fig:TopHiggsxSect}, from which we can see that the highlighted points have a slightly larger production cross section with respect to the SM, and are consistent with the experimental values of the Higgs and top quarks masses along with the experimental and theoretical uncertainties. We observe that modifications of Higgs pair production $\lesssim 20\%$ are possible in this model for our scan results. Plotting the two sensitivity bands corresponding to the CMS and FCC-hh predictions in Figs.~\ref{fig:CMSsensitiv} and~\ref{fig:100TeVsensitiv}, respectively, we see that some parameter points can indeed be excluded through di-Higgs analyses at future collider experiments. Given the relatively small modification of di-Higgs production (which combines with similar observations for single Higgs final states~\cite{Funatsu:2013ni}), a more target approach to constrain this model in the near future is through its lowest lying KK resonances.

\begin{figure}[!t]
 \begin{center}
  \includegraphics[width=1.0\columnwidth]{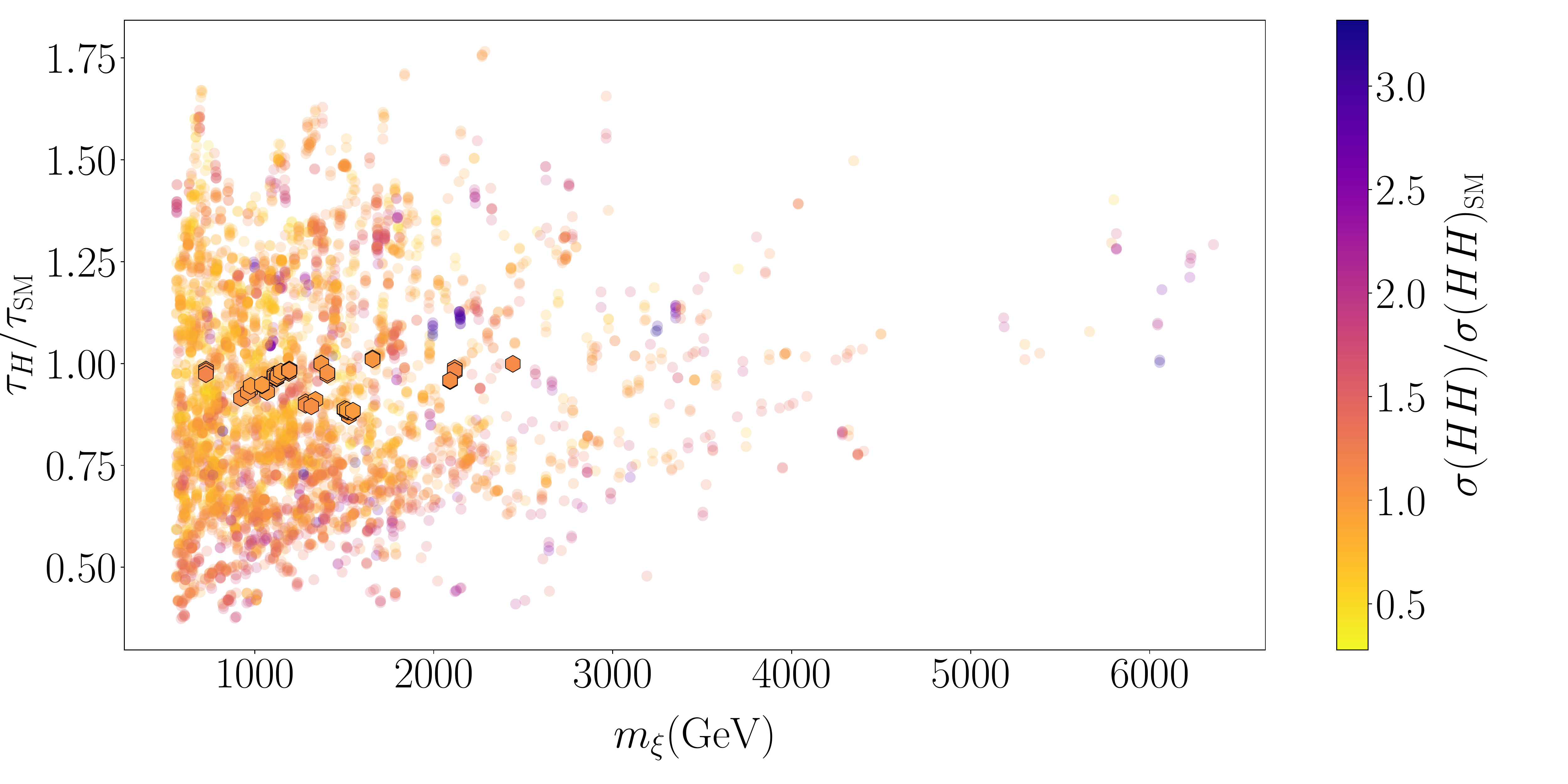}
  \caption{
  Scatter plot correlating exotic mass scale $m_{\xi}$ with the di-Higgs cross section modification. The lowest exotic states are summarised via $m_\xi = \min(m_{\psi^D}, m_{\tau^{(1)}}, m_{b^{(1)}} )$.} \label{fig:LowState}
 \end{center}
\end{figure}

\subsection*{Exotics}
\label{sec:exotics}
\begin{figure}[!b]
 \begin{center}
  \includegraphics[width=1.0\columnwidth]{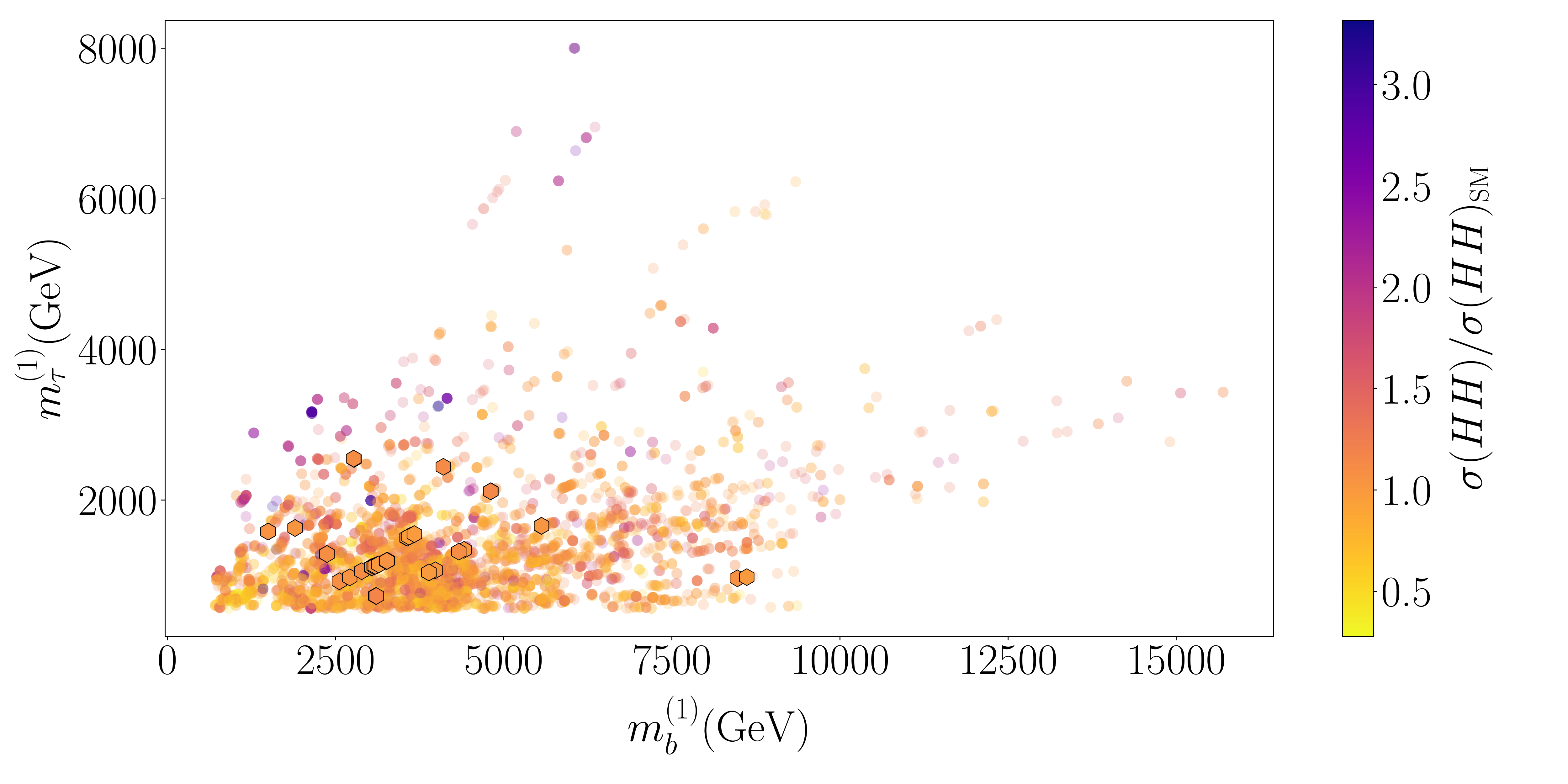}
  \caption{
  Scatter plot correlating the exotic masses for the tau and bottom quark $m_{\tau^{(1)}}, m_{b^{(1)}}$ with the di-Higgs cross section modification.} \label{fig:LowStateBotTau}
 \end{center}
\end{figure}

We now look at the states present in the low energy description that can act as a direct probe of the model. After excluding the points that fell short of the LHC cuts specified in Sec.~\ref{sec:parameter}, we plot the lowest lying exotic mode $m_\xi$ ($\xi=\psi^D, \tau^{(1)}, b^{(1)}$) in Fig.~\ref{fig:LowState}; the lowest lying non-SM modes of the bottom quark, tau lepton and the ``dark fermion'' serve as the next accessible states. We neglect the first excited state of the top quark as it is much heavier than the other exotic states. We can see that most of the viable parameter space points predict that these states lie within the \SI{1}{TeV} to \SI{2}{TeV} range, which should make them accessible by the current colliders via the ongoing searches, which we have highlighted in \ref{sec:parameter}. For the hexagonal points the next accessible state is either the first excitation of the tau lepton or the bottom quark, with the mass correlations plotted in Fig.~\ref{fig:LowStateBotTau}.\footnote{Note that the differential evolution algorithm populates parameter regions that fall outside the LHC analyses that we consider in Sec.~\ref{sec:parameter}, i.e. the fact that these states might be accessible already with data recorded by the LHC experiments does not rule out the model, but would be a sign of additional tuning.} This shows that searches for excited leptons and quarks as they are already pursued by the LHC experiments are crucial tools in further constraining this model.

\section{Summary and Conclusions}
\label{sec:conc}
New physics beyond the Standard Model remains a priority of the current theory and collider phenomenology programme. Efforts split into the study of concrete scenarios as well as more generic approaches to BSM physics using EFT methods. Concrete UV scenarios,
typically contain a vast space of parameters that need to be efficiently sampled to obtain consistent solutions in a fast reliable way.  In this paper we have applied differential evolution to a $SO(11)$ gauge-Higgs unified GUT theory to obtain solutions that are consistent with constraints on the heavy SM states and relevant existing direct collider searches. The efficient way of sampling allows us to widen the UV parameter region, thus considering a more general set of solutions than considered in the literature so far.

As a missing piece of information in the context of this model, we specifically discuss the prospects of di-Higgs production and the model-associated modifications to the inclusive cross section that can be expected. We find that ${\cal{O}}(20\%)$ deviations for the SM can be observed in the light of the constraints that we apply to the model at the TeV scale. This deviation is too small to be a decisive tool in indirectly discovering this model at the LHC, given the latest sensitivity projections provided by CMS~\cite{CMS:2018ccd}. Projections for a potential future FCC-hh~\cite{Contino:2016spe} suggest that some sensitivity can be gained in the di-Higgs modes, however, the most discriminative power lies in the search for the non-SM exotic states. These are loosely bound to the 5D KK scale and thus fall within the capability of a (high luminosity) LHC unless the model is tuned in such a way that the TeV scale becomes vastly separated from the KK scale. On the basis of
our scans we identify the first excitations bottom quark and tau lepton towers as relevant exotic states as ideal candidates for this scenario.

\acknowledgements
C.E. and D.J.M. are supported by the UK Science and Technology Facilities Council (STFC) under grant ST/P000746/1. C.E. acknowledges support by the Durham Institute for Particle Physics Phenomenology (IPPP) Associateship Scheme. D.D.S. is supported by a University of Glasgow College of Science \& Engineering PhD scholarship.


\appendix
\section*{Appendix}
\subsection{KK Tower Equations and Effective potential contributions} \label{appendix:EOMs}
The effective potential $V_\text{eff} (\theta_H)$, and the relevant fields are determined by the KK tower equations which have an explicit $\theta_H$ dependence. To this extent, the bosonic and fermionic sector consists of
\begin{subequations}
\label{eq:tower}
\begin{widetext}
  \begin{equation}
    \begin{cases*}
      2S(1; \lambda_W) C(1 ; \lambda_W) + \lambda_W \sin^2\theta_H  = 0 , \\ \vspace{0.4cm}
      2S(1; \lambda_Z) C(1 ; \lambda_Z) + \dfrac{1}{1- (\sin^2 \theta_W)_\text{EW} } \lambda_Z \sin^2\theta_H  = 0. \label{eqn:BosTower}
    \end{cases*}
  \end{equation}
\begin{equation}
  \begin{cases}
  S_L (1; \lambda_t, c_0) S_R (1; \lambda_t, c_0) + \sin^2 \dfrac{\theta_H}{2} = 0 \\
  S_L(1 ; \lambda_b, c_0) S_R (1; \lambda_b, c_0) + \sin^2 \dfrac{\theta_H}{2} = - \dfrac{ \mu^2_1 S_R(1; \lambda_b, c_0) C_R(1; \lambda_b, c_0) S_L(1; \lambda_b, c_1) C_R(1; \lambda_b, c_1)}{\mu^2_\mathbf{11} (C_R(1; \lambda_b, c_1))^2 - (S_L(1; \lambda_b, c_1))^2} \label{eqn:FermTower}
  \end{cases}
\end{equation}
\begin{equation}
\begin{cases}
  S_L(1 ; \lambda_\tau, c_0) S_R (1; \lambda_\tau, c_0) + \sin^2 \dfrac{\theta_H}{2} = - \dfrac{ \tilde{\mu}^2_2 S_L(1; \lambda_\tau, c_0) C_L(1; \lambda_\tau, c_0) S_R(1; \lambda_b, c_2) C_L(1; \lambda_\tau, c_2)}{\mu'^2_\mathbf{11} (C_L(1; \lambda_\tau, c_2))^2 - (S_R(1; \lambda_\tau, c_2))^2} \\ \\
  - \dfrac{k \lambda_\nu + M}{m_B} \left[ S_L (1; \lambda_\nu, c_0) S_R (1; \lambda_\nu, c_0)  + \sin^2 \dfrac{\theta_H}{2} \right] - \dfrac{m_B}{2k} S_R (1; \lambda_\nu, c_0) C_R (1; \lambda_\nu, c_0) = 0  \label{eqn:LeptonTower}
\end{cases}
\end{equation}
\begin{equation}
S_L (1; \lambda_\psi, c'_0) S_R (1; \lambda_\psi, c'_0) + \cos^2 \dfrac{\theta_H}{2} = 0 \label{eqn:PsiDTower}
\end{equation}
\end{widetext}
\end{subequations}
where the $\lambda_i$ refer to the KK mass eigenstates of the respective fields that are determined through the above system of equations. $\theta_H$ is the value of the Higgs minimum. $S, C$
are boson-related Bessel functions encountered in warped backgrounds evaluated at $z=1$. Similarly $S_L, S_R, C_R, C_L (z, \lambda, c)$ are the fermion-related Bessel functions evaluated at $z=1$ for the various fermionic bulk masses $c$. The other parameters are detailed in Sec.~\ref{sec:parameter}. For the explicit form of the functions see~\cite{Hosotani:2017edv, Furui:2016owe}. The solutions of the system above yields the mass spectra for the various fields as functions of the curvature $m_n(\theta_H) = \lambda_n k$.

The one loop effective potential resulting from the KK tower contributions with mass  is given by,
\begin{equation}
  V_\text{eff}(\theta_H) = \int_{0}^\infty \frac{d^4 p}{(2 \pi)^4} \sum_n  \pm \frac{1}{2} \ln(p^2 + m_n(\theta_H))\,,
\end{equation}
where the $\pm$ sign is related to bosonic/fermionic contributions. The above can be recast by rewriting the tower of Eq.~\eqref{eq:tower} in the form,
\begin{equation}
  1 + \tilde{Q}(\lambda_n) f(\theta_H) = 0\,,
\end{equation}
where we also define $Q(q) = \tilde{Q}(i q z_L^{-1})$. This in turn recasts the contributions in the general form,
\begin{equation*}
   V_\text{eff} (\theta_H) = \pm a \left[ \frac{(k z_L^{-1})}{(4\pi)^2} \int_{0}^\infty dq \,q^3 \ln \left(1 + Q(\lambda_n) f(\theta_H)\right) \right]\,,
\end{equation*}
where $a$ is a field specific constant that accounts for the degrees of freedom. For bosons this also implies a gauge fixing term $\xi$, while for fermions it takes into account the Dirac components of the towers and their colour charges.

To be able to find the mass spectra of the model we need to compute the minimum of the potential $\theta_H$. This is done in via numerical integration of the various contributions.
The contributions are expressed in the $R_\xi = 0$ gauge, and come from all the fields that have 0 modes for both the 5th and 6th dimension and have an explicit $\theta_H$ dependence.

 The effective potential consists of bosonic and fermionic contributions and has the form,
 \begin{align*}
  V^\text{Bosons}_\text{eff} (\theta_H) = & V_\text{eff}^{W^\pm} + V_\text{eff}^{Z^0} + V_\text{eff}^{A^{a, 4}_z, A^{a, 11}_z}  ,\\
  V^\text{Fermions}_\text{eff} (\theta_H) = & V^\text{Top}_\text{eff} + V^\text{Bottom}_\text{eff}  + V^\text{Tau}_\text{eff} + V^\text{Neutrino - 1}_\text{eff} \\ & +  V^\text{Neutrino - 2}_\text{eff} +  V^\text{Dark Multiplet}_\text{eff} .
 \end{align*}
Note that we include the 2nd neutrino sector in the effective potential contribution, but neglect exploring the mass spectrum. For the explicit form of the contributions see~\cite{Hosotani:2017edv}.


\end{document}